\def \SAIT #1 #2 {{\em Mem.\ Soc.\ Astron.\ It.\/} {\bf #1}, #2}
\def \MESS #1 #2 {{\em The Messenger\/} {\bf #1}, #2}
\def \ASTRNACH #1 #2 {{\em Astron. Nach.\/} {\bf #1}, #2}
\def \AAP #1 #2 {{\em Astron. Astrophys.\/} {\bf #1}, #2}
\def \AAL #1 #2 {{\em Astron. Astrophys. Lett.\/} {\bf #1}, L#2}
\def \AAR #1 #2 {{\em Astron. Astrophys. Rev.\/} {\bf #1}, #2}
\def \AAS #1 #2 {{\em Astron. Astrophys. Suppl. Ser.\/} {\bf #1}, #2}
\def \AJ #1 #2 {{\em Astron. J.\/} {\bf #1}, #2}
\def \ANNREV #1 #2 {{\em Ann. Rev. Astron. Astrophys.\/} {\bf #1}, #2}
\def \APJ #1 #2 {{\em Astrophys. J.\/} {\bf #1}, #2}
\def \APJL #1 #2 {{\em Astrophys. J. Lett.\/} {\bf #1}, L#2}
\def \APJS #1 #2 {{\em Astrophys. J. Suppl.\/} {\bf #1}, #2}
\def \APSS #1 #2 {{\em Astrophys. Space Sci.\/} {\bf #1}, #2}
\def \ASR #1 #2 {{\em Adv. Space Res.\/} {\bf #1}, #2}
\def \BAIC #1 #2 {{\em Bull. Astron. Inst. Czechosl.\/} {\bf #1}, #2}
\def \JSQRT #1 #2 {{\em J. Quant. Spectrosc. Radiat. Transfer\/} {\bf #1}, #2}
\def \MN #1 #2 {{\em Mon. Not. R. Astr. Soc.\/} {\bf #1}, #2}
\def \MEM #1 #2 {{\em Mem. R. Astr. Soc.\/} {\bf #1}, #2}
\def \PLR #1 #2 {{\em Phys. Lett. Rev.\/} {\bf #1}, #2}
\def \PASJ #1 #2 {{\em Publ. Astron. Soc. Japan\/} {\bf #1}, #2}
\def \PASP #1 #2 {{\em Publ. Astr. Soc. Pacific\/} {\bf #1}, #2}
\def \NAT #1 #2 {{\em Nature\/} {\bf #1}, #2}

\documentstyle[psfig]{memsait}
\input epsf.sty
\begin{opening}
\title{MOLECULES IN QSOS AND QSO ABSORPTION LINE SYSTEMS
AT HIGH REDSHIFT} 
\author{Patrick Petitjean}
\institute{Institut d'Astrophysique de Paris, 98bis Bd Arago,
75014 Paris, France\\
DAEC, Observatoire de Paris, 5 Place J. Janssen, 92195 Meudon, France}
\date{} 
\end{opening}

\begin{document}

\oddpagefooter{}{}{} 
\evenpagefooter{}{}{} 
\ 
\bigskip

\begin{abstract}
Molecules dominate the cooling function of neutral metal-poor gas at high 
density. Observation of molecules at high redshift is thus 
an important tool toward understanding the physical conditions
prevailing in collapsing gas.
Up to now, detections are sparse because of small filling factor 
and/or sensitivity 
limitations. However, we are at an exciting time where new capabilities 
offer the propect of a systematic search either in absorption using the
UV Lyman-Werner H$_2$ bands or in emission using the CO emission lines 
redshifted in the sub-millimeter.
\end{abstract}
\section{Damped Ly$\alpha$ systems}
\subsection{Introduction}
QSO absorption line systems probe the baryonic matter over most of the
history of the Universe (0~$<$~$z$~$<$~5).  The so-called damped
Ly$\alpha$ (hereafter DLA) systems are characterized by a very large
H~{\sc i} column density ($N$(H~{\sc i})~$>$~2$\times$10$^{20}$
~cm$^{-2}$), similar to what is usually observed through local spiral disks. 
The case for these systems to be produced by proto-galactic disks is
supported by the fact that the cosmological density of gas associated
with these systems is of the same order of magnitude as the
cosmological density of stars at present epochs (Wolfe 1996). 
The presence of heavy elements ($Z \sim 1/10 ~ Z_\odot$) and
the redshift evolution of metallicity suggest ongoing star
formation activities in these systems (Lu et al. 1996, Pettini et al.
1996, 1997). Moreover,
strong metal line systems have been demonstrated to be associated with
galaxies at low and intermediate $z$ (e.g.  Bergeron \& Boiss\'e 1991).
It has also been shown that the profiles of the lines arising in the
neutral gas show evidence for rotation (Wolfe 1996, Prochaska \& Wolfe
1997). Whether these arguments are enough to demonstrate that DLA
systems arise in large disks is a matter of debate however. Indeed
simulations have shown that the progenitors of present day disks of
galaxies could look like an aggregate of well separated dense clumps at
high redshift. The kinematics could be explained by relative motions of
the clumps with very little rotation (Haehnelt et al. 1997, Ledoux et al.
1998). Moreover,
using {\sl HST} high spatial resolution images of the field of seven
quasars whose spectra contain DLA lines at intermediate redshifts
(0.4~$<$~$z$~$<$~1), Le~Brun et al. (1997) show that, in all cases,
at least one galaxy candidate is present within 4~arcsec from the
quasar. There is no dominant
morphological type in their sample: three candidates are spiral
galaxies, three are compact objects and two are amorphous low surface
brightness galaxies. Therefore, although the nature of the DLA systems is
unclear they trace the densest regions of the Universe where
star formation occurs.\par
\subsection{Molecular hydrogen}
\begin{figure}
\epsfysize=15cm 
\epsfxsize=11cm 
\hspace{1.5cm}\epsfbox{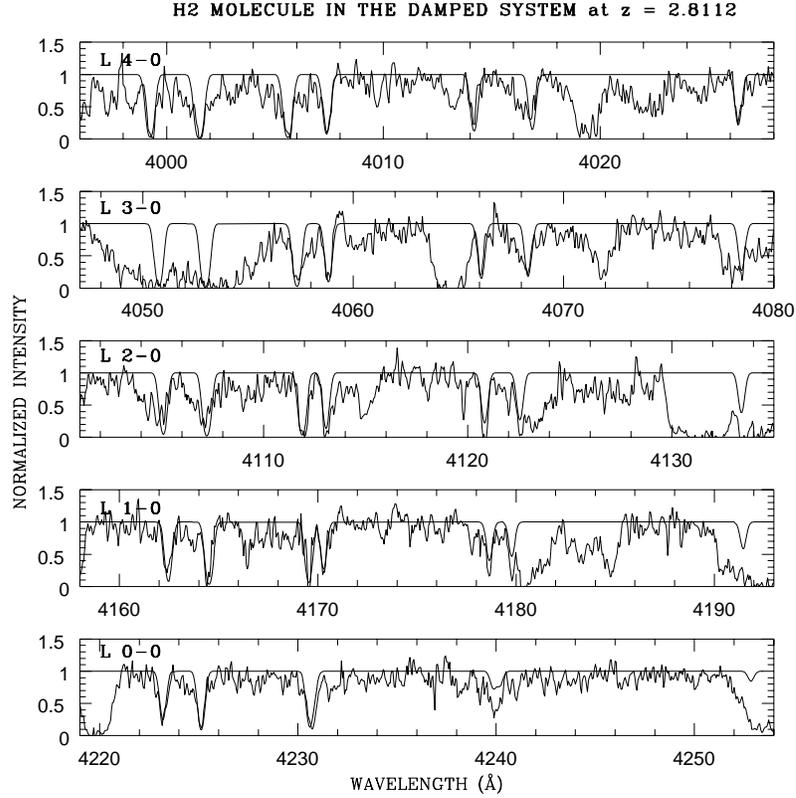} 
\caption[h]{Fit result for a few rotational transitions of
the H$_2$ Lyman absorption bands in the $z_{\rm abs}$~=~2.8112 
system toward PKS~0528--250. The spectrum has been obtained
with the echelle spectrograph CASPEC attached on the ESO 3.6~m
at La Silla. The resolution is $R$~=~36000 and the integration
time 5~hours.}
\end{figure}
It is thus surprising that despite intensive searches, the amount of
H$_2$ molecules seems quite low in damped Ly$\alpha$ systems in
contrast to what is observed in our own galaxy. Two detections of H$_2$
molecules in high redshift DLA systems have been reported.  Recently Ge
\& Bechtold (1997) have found strong absorptions in the $z_{\rm
abs}$~=~1.9731 DLA system toward Q~0013--004. They derive
$N$(H$_2$)~=~6.9$\times$ 10$^{19}$~cm$^{-2}$, $b$~=~15~km~s$^{-1}$,
$T_{\rm ex}$~$\sim$~70~K and $n$(H)~$\sim$~300~cm$^{-3}$ for a total
hydrogen column density $N$(H)~=~6.4$\times$10$^{20}$~cm$^{-2}$. This
system has by far the largest H$_2$ abundance
$f$~=~2$N$(H$_2$)/[2$N$(H$_2$)~+~$N$(H~{\sc i})] $\sim$~0.22$\pm$0.05
observed in DLA systems. However the exact number should be confirmed 
using higher resolution data. Other searches have led to much smaller
values or upper limits ($f$~$<$~10$^{-6}$, Black et al. 1987, Chaffee
et al. 1988, Levshakov et al. 1992). 
Table~1 summarizes the caracteristics of damped Ly$\alpha$ systems 
that have been searched for molecules.

Levshakov \& Varshalovich (1985)
suggested that molecules could be present toward PKS~0528--250 
at a redshift ($z_{\rm abs}$~=~2.8112), slighly larger than
the emission redshift of the quasar. This claim has been confirmed by Foltz et
al. (1988) using a 1~\AA~ resolution spectrum. The latter authors
derive $N$(H$_2$)~=~10$^{18}$~cm$^{-2}$, $b$~=~5~km~s$^{-1}$, $T_{\rm
ex}$~=~100~K and log~$N$(H~{\sc i})~=~21.1$\pm$0.3. By fitting the
damped absorption together with the Ly$\alpha$ emission from the
quasar, M\o ller \& Warren (1993) find log~$N$(H~{\sc i})~=~21.35.

New high resolution data has been recently obtained by Srianand \& Petitjean (1998).
They estimate the column density of H$_2$ molecules 
$N$(H$_2$)~$\sim$~6$\times$10$^{16}$~cm$^{-2}$ and the fractional 
abundance of H$_2$, $f$~=~5.4$\times$10$^{-5}$ (see Fig.~1). 
The excitation
temperature derived for different transitions suggests that the
kinetic temperature of the cloud is $\sim$200~K
and the density $n$~$\sim$~1000~cm$^{-3}$. The cloud has therefore
a dimension along the line of sight smaller than 1~pc. Since
it obscurs the broad-line emission region, its transverse dimension should
be larger than 10~pc.

Upper limits are obtained on the column densities of C~{\sc i} 
($<$~10$^{12.7}$~cm$^{-2}$)
and CO ($<$~10$^{13.2}$~cm$^{-2}$; $N$(CO)/$N$(H~{\sc i})~$<$~7$\times$10$^{-9}$). 
It is suggested that the
ratio $N$(H$_2$)/$N$(C~{\sc i}) is a useful indicator of the 
physical conditions in the absorber. 
Photo-ionization models 
show that radiation fields with spectra similar to typical AGNs or
starbursts are unable to reproduce all the constraints and in particular
the surprizingly small $N$(C~{\sc i})/$N$(H$_2$) and
$N$(Mg~{\sc i})/$N$(H$_2$) ratios. 
In view of the models explored, the most likely ionizing spectrum
is a composite of a UV-"big bump" possibly produced by a local starburst
and a power-law spectrum from the QSO that provides the X-rays.
This suggests that the gas is not predominantly
ionized by the quasar and that star-formation may occur in the clouds,
a conclusion reached as well by Warren
\& M\o ller (1996) and Ge et al. (1997). 
Dust is needed to explain the production of molecules in the cloud. The
amount of dust is broadly consistent with the [Cr/Zn] abundance
determination. 
%

\par\noindent
\vspace{0.5cm} 
\par\noindent
\centerline{\bf Tab. 1 - H$_2$ molecules in DLA systems}
\begin{table}[h]
\hspace{1.5cm} 
\begin{tabular}{|l|c|c|c|c|c|c|}
\hline
Name           & 000-263 & 0013-004 & 0100+130  & 0528-250 & 1331+170 &
1337+113 \\
\hline
$z_{\rm em}$   & 4.110 & 2.084  & 2.681  & 2.770  & 2.081 &  2.919\\
$z_{\rm abs}$  & 3.391 & 1.9731 & 2.309  & 2.811  & 1.776 &  2.796\\
$N$(HI) (10$^{21}$~cm$^{-2}$ & 2.0 & 0.64  & 2.5  & 2.2  &  1.5  & 0.80\\
$N$(H$_2$) (10$^{16}$~cm$^{-2}$& $<$~0.3  & 6900 & $<$~0.5 &  6 & ...  & $<$~5\\
$f_{{\rm H}2}$(10$^{-4}$)  & $<$~0.03  & 2200 & $<$~0.04 & 0.5  & ... &
$<$~1.3\\
\hline
\multicolumn{7}{l}{Levshakov et al. (1992); Ge \& Bechtold (1997); Srianand
\& Petitjean (1998)} \\
\end{tabular}
\label{tsys}
\end{table}
%
%
\begin{figure}
\centerline{
\psfig{figure=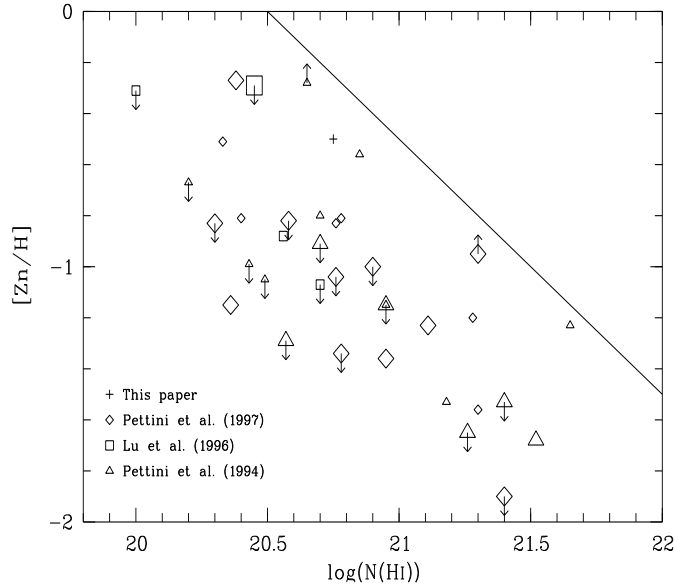,width=9.cm,height=8.cm,angle=270}
}
\caption[h]{
[Zn/H] versus log~$N$(H~{\sc i}) for 37 damped Ly$\alpha$ systems; 
small symbols correspond to $z_{\rm abs}$~$<$~2.15 and large symbols to 
$z_{\rm abs}$~$>$~2.15.
The line in the upper right corresponds to $N$(Zn~{\sc ii}) = 
1.4$\times$10$^{13}$~cm$^{-2}$ or 
to Galactic material inducing $A_{\rm V}$~$\sim$~0.27. The figure is taken
from Boiss\'e et al. (1998).}
\label{boisse}
\end{figure}
\subsection{Is there a bias against detection of H$_2$ molecules ?}
The small number of H$_2$ detections in damped systems is intriguing. 
Indeed in the interstellar medium of our Galaxy,
all the clouds with log~$N$(H~{\sc i})~$>$~21 have 
log~$N$(H$_2$)~$>$~19 (Jenkins \& Shaya 1979). 
Formation of H$_2$ is expected on the surface of dust grains if the gas
is cool, dense and mostly neutral, and from the formation of negative
hydrogen if the gas is warm and dust free (see e.g. Jenkins \& Peimbert
1997). Destruction is mainly due to UV photons. The effective
photodissociation of H$_2$ takes place in the energy range 11.1--13.6
eV, through Lyman-Werner band line absorption. 

In the DLA system toward PKS~0528-250,
(i) abundances are of the order of 0.1~$Z_{\odot}$;
(ii) the ratio [Cr/Zn] indicates a depletion factor into dust-grains
of the order of half of that in the Galactic ISM; 
(iii) although it has been shown that the cloud is located at a distance 
larger than 
10~kpc from the quasar, it is still close to it and exposed to its UV flux.
Nonetheless, molecular hydrogen is detected. This suggests that indeed,
molecular hydrogen should be seen in most of the damped systems.
The small number of detections may be explained if observations are biased 
against the presence of molecules. Indeed it can be speculated that molecules 
should be found 
predominantly in gas with a non negligible amount of dust.
However the corresponding extinction of the background quasar due to the dust 
in the damped system could be large enough to
drop the quasar out of the sample of quasars that are usually observed 
for such studies.
Boiss\'e et al. (1998) notice that for the damped systems studied up
to now, the larger the H~{\sc i} column density, the smaller the 
abundances (see Fig.~\ref{boisse}). This suggests that 
the high column density DLA systems 
detected up to now are those with the smallest metallicities and 
consequently those with the smallest amount of dust. 
One way to clear up this problem is to observe a complete sample of quasars
(if possible constructed without color selection) and to search for high 
column density damped systems.  
\section{CO emission in QSOs at high redshift}
%
%
\begin{figure}
\vbox{\centerline{
\psfig{figure=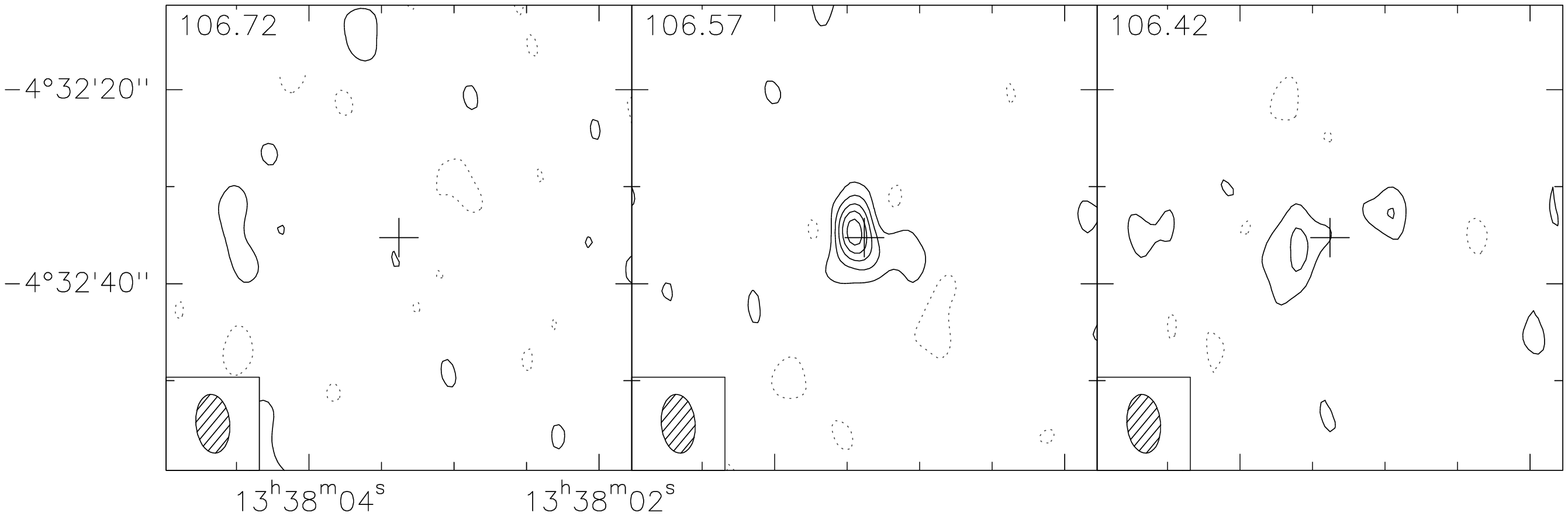,width=14.cm,height=4.5cm,angle=270}}
\centerline{
\psfig{figure=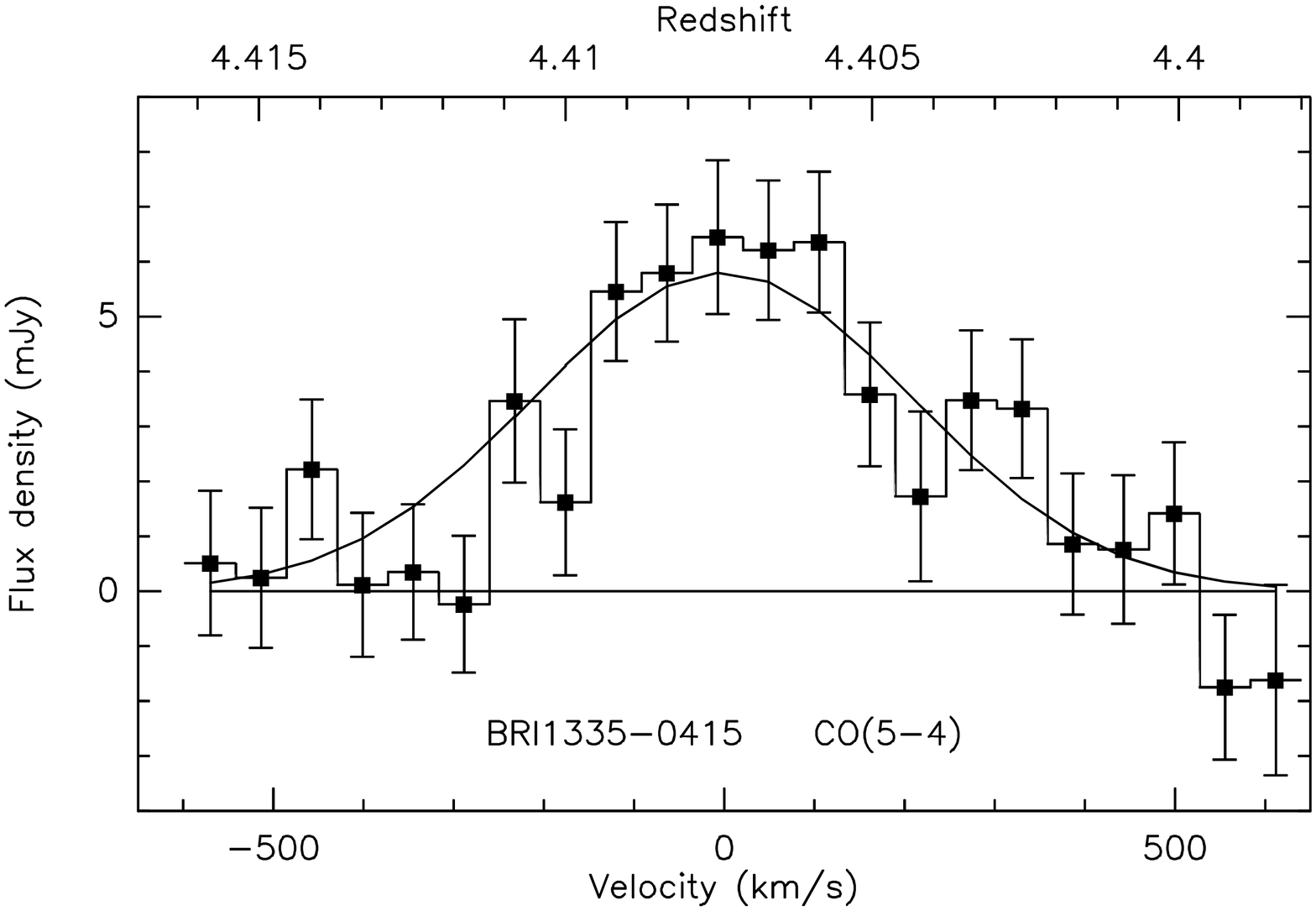,width=8.cm,height=6.5cm,angle=270}}
}
\caption[h]{Upper panel: Channel maps of the CO J=5-4 line towards 
BRI~1335-0415 (Guilloteau et al. 1997).
Channel width is 150 MHz, i.e. 420 km~s$^{-1}$. The contour step is
0.8 mJy/beam (2 $\sigma$). The observing frequency in GHz is indicated in the
upper left corner of each map.\par\noindent
Lower Panel: Spectrum of the CO J=5-4 line towards BRI~1335-0415, superimposed
with the best Gaussian profile. Errorbars are $\pm 1 \sigma$.
The velocity scale corresponds to a frequency of 106.570 GHz corresponding
to a redshift $z=4.4074\pm0.0015$.
}
\label{q1335}
\end{figure}
%

%

The initial detection of CO in FIRAS10214+4724 at $z=2.3$ (Brown \&
Vanden Bout 1992, Solomon et al. 1992) has been followed by CO detections in
H1413+117 (The Cloverleaf) at $z=2.5$ (Barvainis et al. 1994),
and more recently, in 53W002 at $z=2.34$ (Scoville et al. 1997). In the latter object,
CO(3--2) is detected using the OVRO interferometer. 
This object is interesting because it appears to lie within a
cluster of roughly 20 Ly$\alpha$ emission line objects, the most
distant such cluster known (Pascarelle et al 1996).
In addition, CO(3-2) line emission is detected from the gravitationally lensed quasar 
MG~0414+0534 at redshift 2.64 (Barvainis et al. 1998), using the
IRAM Plateau de Bure Interferometer. The line is broad, 
with $\Delta$v$_{FWHM}$~= 580~km~s$^{-1}$. The velocity-integrated CO flux is
comparable to, but somewhat smaller than, that of IRAS F10214+4724 and the 
Cloverleaf quasar (H1413+117), both of which are at similar redshifts. 

It is worth noting that all of the CO-detected objects mentioned above
have also been detected in the far-IR (IRAS) or 
1~mm/submm continuum. This 
indicates that these objects also contain large amounts of dust.
Indeed, thermal emission has now been convincingly detected at 1.3\,mm in
$\sim$8 $z>2$ QSOs (Barvainis et al. 1992; McMahon et al. 1994; Isaak
et al. 1994; Ivison 1995; Omont et al 1996b) and in one radio-galaxy at
$z > 2$ (Dunlop et al. 1992; Chini \& Kr\"{u}gel 1994; Hughes et al.
1997). The strategy thus 
adopted to detect CO in high-redshift quasars is 
(i) to search for dust emission toward radio-quiet quasars 
in the millimeter or submillimeter ranges using
single dish telescopes (Omont et al. 1996b); 
(ii) to follow-up the detections:
search for CO emission and mapping the continuum using the IRAM
interferometer (Omont et al. 1996a; Guilloteau et al. 1997).

\begin{figure}
\centerline{
\psfig{figure=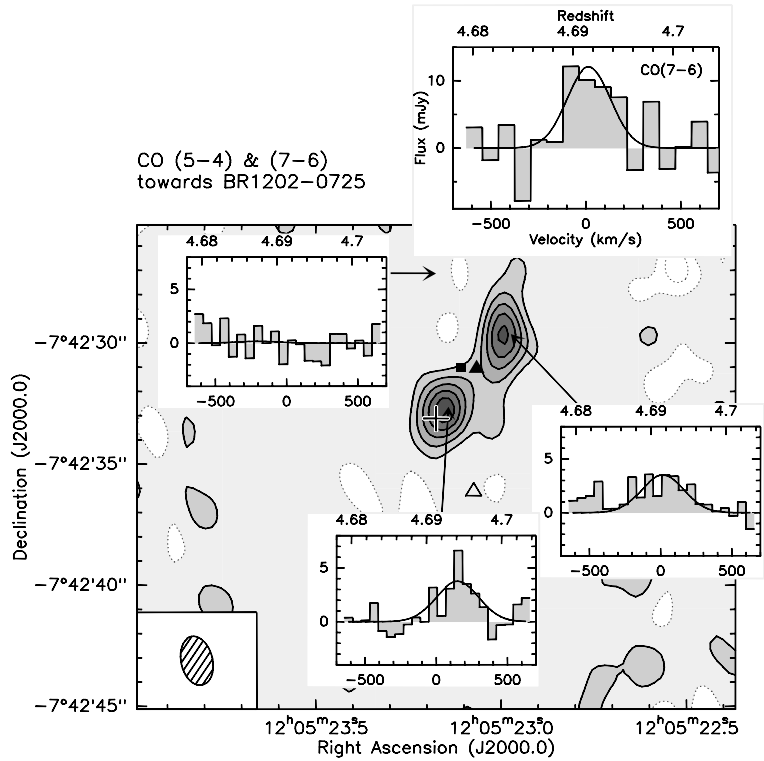,width=10.cm,height=10.cm,angle=270}
}
\caption[h]{Image of the field surrounding the radio-quiet 
quasar BR1202--0725 at redshift $z_{\rm em}$~=~4.7 taken in the 1.35~mm continuum
with the IRAM interferometer of the Plateau de Bures (Omont et al. 1996a).  
Superposed are the positions of the different visible features:~
Cross: QSO; Filled Triangle: Continuum feature; Filled Square: Ly$\alpha$
peak; Open Triangle : Second continuum feature. 
Two peaks of emission are seen, one centered on the quasar, the second
4~arcsec in the north--west direction. 
Also shown are the spectra of the CO(5-4) line emission at the position of
the two continuum peaks and the
spectrum of CO(7--6) line observed with the
IRAM 30m telescope. 
}
\label{q1202}
\end{figure}

The two strongest 1.3~mm continuum emitters, BR~1202-0725 
($z_{\rm em}$~$\sim$~4.7)
and BRI~1335-0415 ($z_{\rm em}$~$\sim$~4.4) have been detected 
in CO(5-4) with comparable total integrated line intensities
(2.4$\pm$0.5, see Fig.~\ref{q1202}, and 2.8$\pm$0.3~Jy~km~s$^{-1}$,
see Fig.~\ref{q1335}, respectively).
The corresponding mass of molecular gas is of the order of 
$\sim 10^9 - 10^{11} M_{\odot}$.

Such a jump in the redshift range of millimeter radioastronomy was
made possible by: (i) the existence at large redshift of
objects with molecular gas and dust contents on the one hand
and starburst activity on the other 
similar to what is observed in the most powerful ultra-luminous IRAS galaxies; 
(ii) the increase with redshift of the apparent luminosity 
of a dusty object observed
at fixed wavelength; (iii) the existence of gravitational
lensing amplification at least in FIRAS10214+4724
and H1413+117. However, since some tentative detections of quasars
and quasar absorption line systems have not been confirmed (Wiklind \&
Combes 1994; Braine, Downes \& Guilloteau 1996), the number of
detections of CO at $z>2$ remains small. 

The detection of CO and dust at millimeter wavelength in a few
prominent objects at very large redshift opens the possibility to
study the molecular gas in high redshift galaxies. 
Although at present detections are biased toward 
highly luminous objects with intense 
activity (either starbursts or AGN or both), they open the exciting prospect 
to study more standard objects with the new instrumentation 
planned to be built soon.

%
\acknowledgements
The results presented in this papers have been obtained in
collaboration with Alain Omont, St\'ephane Guilloteau and
R. Srianand. I would like to thank the organisers of the meeting for their
hospitality and kindness.


\end{document}